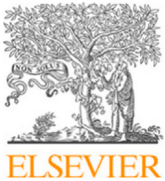
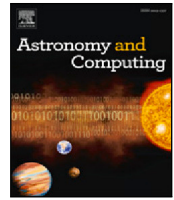

Full length article

# The GRBSN webtool: An open-source repository for gamma-ray burst-supernova associations

Gabriel Finneran ⓘ *, Laura Cotter ⓘ, Antonio Martin-Carrillo ⓘ

*School of Physics and Centre for Space Research, University College Dublin, Dublin, D04 V1W8, Dublin, Ireland*

ABSTRACT

This paper presents the GRBSN webtool, an open-source data repository coupled to a web interface that hosts the most complete dataset of GRB-SN associations to date. In contrast to repositories of supernova (SN) or gamma-ray burst (GRB) data, this tool provides a multi-wavelength view of each GRB-SN association. GRBSN allows users to view and interact with plots of the data; search and filter the whole database; and download radio, X-ray, optical/NIR photometric and spectroscopic data related to a GRB-SN association. The web interface code and GRB-SN data are hosted on a public GitHub repository, allowing users to upload their own data, flag missing data and suggest improvements. The GRBSN webtool will be maintained by the Space Science group at University College Dublin, Ireland. As the number of confirmed GRB-SN associations increases in the coming years, the GRBSN webtool will provide a robust framework in which to catalogue these associations and their associated data. The web interface is available at: https://grbsn.watchertelescope.ie.

## 1. Introduction

Gamma-ray bursts (GRBs) are brief, intense flashes of gamma-ray radiation which result from the death of a massive star or the merger of compact objects. GRBs are detected daily (Gehrels and Mészáros, 2012) by space-based high-energy detectors, such as those on-board the *Swift* (Gehrels et al., 2004), *Fermi* (Meegan et al., 2009) and *SVOM* (Atteia et al., 2022) satellites. GRBs may be categorised into long and short bursts based on the duration of their gamma-ray emission (Kouveliotou et al., 1993). It was suggested that the long-GRBs (>2s) were associated with the collapse of massive stars (collapsar type GRBs), while the short-GRBs were thought to be due to the merging of compact objects such as neutron stars (merger type GRBs). However, in recent years the discovery of long-mergers (e.g. GRB211211A Troja et al., 2022) and short-collapsars (e.g. GRB200826A Ahumada et al., 2021) has blurred this definition significantly.

In the *Fireball Model*, GRBs are powered by the liberation of energy from an ultra-relativistic jet ($\Gamma$ ~100) launched and powered by the *central-engine* of the event (Rees and Meszaros, 1994). In the case of collapsar GRBs, the central engine is likely an accreting black hole (Woosley, 1993), while for merger GRBs, a spinning-down highly magnetised neutron star may act as the central engine (Usov, 1992) (see Salafia and Ghirlanda, 2022 for a recent review of jet-launching in these events).

The ultra-relativistic jet is responsible for the characteristic emission episodes of a GRB: *prompt* and *afterglow* emission. In the fireball model, prompt emission is generated within the GRB jet due to the liberation of the jet energy at internal shocks (Daigne and Mochkovitch, 1998; Rees and Meszaros, 1994). These internal shocks form when shells of material with differing velocities launched by the central engine collide, forming the internal shocks and releasing energy by synchrotron and thermal emission (Daigne and Mochkovitch, 1998). The afterglow is created when the jet decelerates due to its interaction with the surrounding medium, producing shocks which liberate energy in the form of broadband synchrotron emission, visible from X-rays to radio wavelengths (Mészáros and Rees, 1997; Rees and Meszaros, 1994, 1992). An extensive overview of GRBs can be found in Zhang (2018).

Direct evidence for the link between GRBs and massive star death came in April 1998, with the discovery of supernova (SN) SN1998bw at the same coordinates as GRB980425 just days after the GRB was detected (Galama et al., 1998). The subsequent detection of GRB030329-SN2003dh (Hjorth et al., 2003) confirmed the link between SNe and GRBs. To date, more than 60 *GRB-SN associations* have been found (for a review of GRB-SN associations see Cano et al., 2017). In the majority of cases, the SNe associated with GRBs are broad-line type Ic (Ic-BL) SNe.[1] Spectroscopically these SNe are identified by the absence of hydrogen and helium absorption features, and broad absorption

---

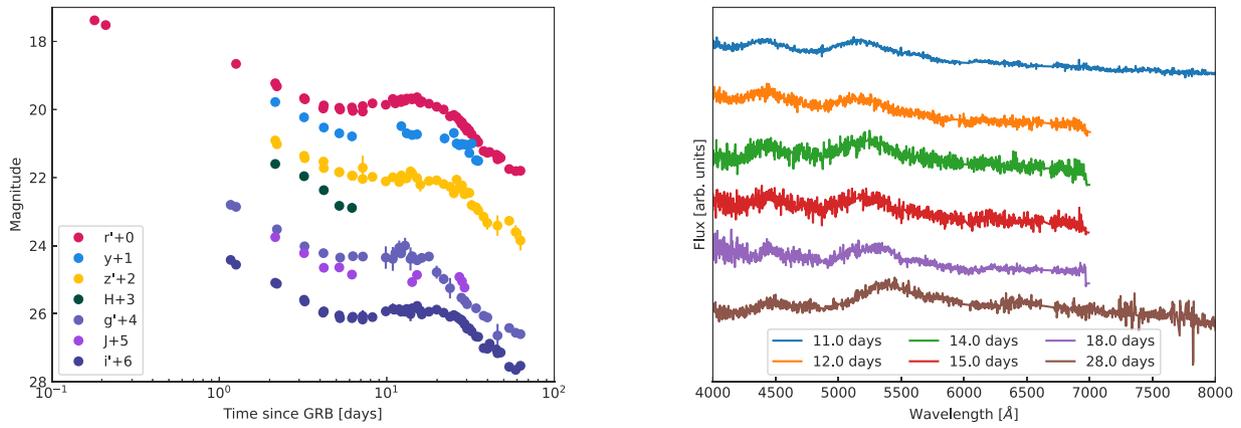

**Fig. 1.** *Left:* Optical lightcurves of GRB130702A-SN2013dx, showing the typical early decline of the GRB component followed by the SN peak at 10–15 days. Optical data reproduced from Toy et al. (2016). *Right:* Spectral sequence of GRB130702A-SN2013dx, showing the typical broad features of a Ic-BL supernova. Spectra reproduced from D'Elia et al. (2015). For both figures the times are relative to the GRB trigger-time of 2013-07-02T00:05:23.079 (Cheung et al., 2013).

features of Fe II, Si II, Ca II and O I (Ho, 2020). They also tend to have much higher velocities than Ic supernovae (Modjaz et al., 2016), sometimes up to 60000 km/s (e.g. SN2020bvc Rho et al., 2021). Photometrically Ic-BL SNe tend to peak at an absolute $r$-band magnitude of $-18.6$ (Taddia et al., 2019), exhibiting a rapid rise and peak at 10–15 days, followed by a decline over several weeks. Fig. 1 shows the optical lightcurves and spectral sequence of GRB130427A-SN2013dk taken from the GRBSN webtool.

Multiwavelength observations are critical to the study of a GRB-SN association. Detection of the GRB high-energy prompt and afterglow components relies on $\gamma$/X-ray observations from telescopes such as *Swift* (Gehrels et al., 2004). On the other hand, supernovae are usually detected by large optical/near-infrared (NIR) survey telescopes, such as the *Zwicky Transient Facility* (ZTF) (Bellm et al., 2019) or the *Asteroid Terrestrial Impact Last-Alert System* ATLAS (Tonry et al., 2018). Spectroscopic classifications of newly discovered SNe are typically reported to the Transient Name Server (TNS).[2] In the case of a GRB with no prompt emission, commonly known as *orphan afterglows* (off-axis GRBs, choked GRBs or dirty fireballs), the supernova emission may precede the GRB emission (e.g. Izzo et al., 2020). For these events, the first detection is expected to happen in the optical/NIR energy range. In these cases, follow up observations with X-ray instruments may then be necessary to confirm the presence of the GRB afterglow component. At later times (~ yrs) radio observations may be necessary to determine whether a supernova harbours an ultra-relativistic jet associated with an orphan afterglow (e.g. Soderberg et al., 2006; Corsi et al., 2016).

As shown in Fig. 2, there are around one to two confirmed GRB-SN events per year, with some small fluctuations from year to year. This rate included both *spectroscopic* GRB-SNe (those for which a spectroscopic confirmation of the SN was made), and *photometric* GRB-SNe (which are identified by the characteristic bump in lightcurve between 10–15 days post-burst, which can be seen in Fig. 1). The use of large scale survey telescopes has made it easier than ever to detect new Ic-BL supernovae, with a large increase in detections since ZTF began operations in 2018. With future facilities, such as the *Vera Rubin Observatory* (VRO) (Ivezić et al., 2019), surveying large sections of the sky each night, the detection rate of Ic-BL supernovae is expected to increase significantly. This is expected to lead to a large number of new GRB-SN events, including the potential for orphan afterglows. A handful of candidate orphan afterglows have been detected in the past by wide-field surveys, including 6 observed by ZTF through dedicated afterglow searches (Srinivasaragavan et al., 2025). This corresponds to

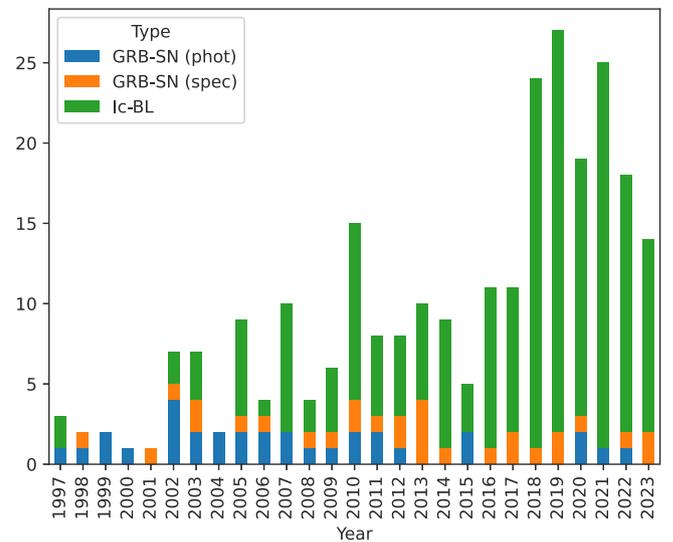

**Fig. 2.** The Ic-BL classification rate has increased significantly over the past 27 years, particularly after 2018 with the advent of ZTF. In contrast, the rate of GRB-SN detections has remained relatively constant over this time. GRB-SNe are marked in this plot where a photometric bump at 10–15 days was observed (GRB-SN (phot)), or where spectroscopic confirmation of the SN was made (GRB-SN (spec)). Sources: WISeREP, the GRBSN webtool and TNS.

< 0.5 orphan afterglow candidates per year from ZTF. In a single visit, ZTF has a typical sensitivity of 20.5 magnitudes (Perley et al., 2024), VRO will enable depths of up to magnitude 24.5 in a single visit (Ivezić et al., 2019), which should lead to an increased number of orphan candidates. More events means more data to be gathered and analysed and it is vital that these data are rapidly and easily accessible for the community.

Existing repositories and webtools are not designed to meet the needs of GRB-SN astronomers. There are currently no tools which display all of the data for every GRB-SN association in one place and in a consistent way. In particular, though some tools such as the *Open Supernova Catalogue*[3] (Guillochon et al., 2017) provide the optical and spectroscopic data for the events which were classified as Ic-BLs, this represents only half the sample, as many GRB-SN do not have a spectroscopically confirmed supernova. Furthermore there is usually no X-ray or radio data available on these tools, both of which are crucial for studying the GRB component. To access the full dataset, researchers must consult multiple sources. Retrieving the data in this way is a long and tedious process that results in duplication of effort. In some cases this data is only available from the original paper and often not in a machine readable format. In some cases these papers are locked behind a paywall, meaning the data may be unavailable to some researchers and to the public.

---

[2] https://www.wis-tns.org

[3] The public facing side of the OpenSN tool is no longer available as of March 8th 2022. The API remains available.

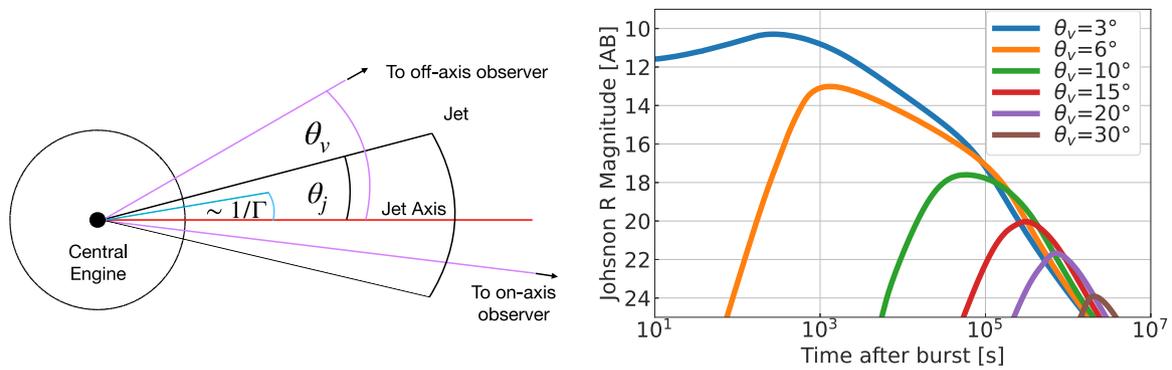

**Fig. 3.** *Left:* Geometry of a GRB, showing the viewing angle $\theta_v$, jet half-opening angle $\theta_j$ and the beaming angle $\sim 1/\Gamma$. *Right:* Simulations of a GRB afterglow showing the influence of the viewing angle on the evolution and morphology of the lightcurve. Simulations were performed using the `afterglowpy` *Python* package (Ryan et al., 2020). The jet half opening angle is $\theta_j$=5 degrees for the top-hat jet. The isotropic energy is $E_{iso} = 10^{52}$ ergs, the redshift is z=0.05, and typical values are used for the other microphysical parameters. Identical values are used in all simulations.

The GRBSN webtool has been developed to solve this problem. The webtool's database includes multiwavelength lightcurves as well as the bulk parameters of both the supernova and gamma-ray burst components. The catalogue currently consists of 61 GRB-SNe, making it the most comprehensive list of GRB-SN associations available to date. These events can be divided into 3 categories: 29 GRBs with a spectroscopically confirmed supernova ("Spectroscopic GRB-SN"), 31 GRBs with a photometrically confirmed supernova ("Photometric GRB-SN") and 1 off-axis GRB detected through its association with a Ic-BL supernova ("Orphan GRB-SN"). The GRBSN webtool will be maintained by the Space Science group at University College Dublin, Ireland. This includes review and standardisation of data submitted to the tool by its users.

Although much has been learned in the past 26 years, there remain many open questions in GRB-SN science, including: what is the relationship between the GRB and SN energy, do they share a common reservoir?; what gives rise to the high velocities in the supernovae associated with GRBs, is it related to jet activity?; how common are orphan GRBs due to choked jets or viewing angle effects?; and what makes the progenitors of these events capable of producing a supernova and a GRB when many stars cannot?. With greater numbers of GRB-SNe expected to be discovered in the coming years, this tool has the potential to make the tracking and investigation of GRB-SN associations much simpler.

This paper presents a detailed description of the GRBSN webtool. The motivation behind the development of this tool is discussed in Section 2. Section 3 presents the frontend of the webtool, while the backend, data collection and standardisation are described in Section 4. Section 4 also discusses methods by which users of the tool may contribute their own data. An application of the webtool to the study of the Amati relation is presented in Section 5 and future development and maintenance is discussed in Section 6.

## 2. Motivation

### 2.1. Data requirements for GRB-SN science

GRB-SNe are visible across the electromagnetic spectrum, and a multi-wavelength dataset is often needed to disentangle the effects of the GRB from those of the SN (e.g. Izzo et al., 2020) and constrain the physical parameters of a GRB-SN. The use of a broad multi-wavelength dataset, spanning from radio to X-rays, is quite common for studies of GRBs, but less common among studies of non-GRB-associated SNe. For this reason, online repositories for supernova data, such as TNS, the Open Supernova Catalogue or WISeREP[4] only show the optical and spectral data for an SN.

The supernova component of a GRB-SN is powered by the decay of $^{56}$Ni and $^{56}$Co (Arnett, 1982), which produces a lightcurve in the optical range which peaks at around 10–15 days. Optical photometry and spectra are used in the calculation of the SN bulk parameters: ejecta mass ($M_{ej}$); nickel mass ($M_{Ni}$); and the kinetic energy of the ejecta ($E_k$) (e.g. Cano et al., 2017). The supernova spectrum is also used to determine the supernova type. The spectral evolution of the SN may also be able to constrain any anisotropies in the SN emission caused by a relativistic jet, which could be especially useful for orphan GRBs (Barnes et al., 2018). In the absence of a spectrum, the optical lightcurve of the GRB may depart from the usual power-law decline of the afterglow after a few days, revealing an underlying SN component that peaks at around 10–15 days after the GRB (e.g. Izzo et al., 2020).

Fig. 3 demonstrates that presence of the ultra-relativistic jet in GRBs gives rise to a viewing angle dependence in the lightcurves of GRBs. Initially, emission from the jet plasma is relativistically beamed, due to the rapid motion of the jet plasma. Radiation is beamed into a cone of opening angle $\theta_j \sim 1/\Gamma$, where $\Gamma$ is the bulk Lorentz factor of the jet (Rhoads, 1997). In the case of an on-axis GRB, when the observer's line of sight is within the beaming cone, prompt GRB emission is detected immediately. Following this, the afterglow emission dominates the lightcurve at early times (<1 day). However, when a GRB is accompanied by a supernova, Fig. 1 shows that the SN emission becomes visible ~5 days after the initial explosion, limiting our ability to study the jet properties at these late times. It is in this case that the X-ray monitoring of GRB-SNe becomes extremely important. As the emission in X-rays is mainly due to the afterglow of the GRB, generated via synchrotron emission at the forward shock of the GRB jet, the X-ray emission probes the physics of the GRB and its environment without contamination from the SN at any time.

---
[4] https://www.wiserep.org

In the case of an off-axis GRB, when the viewing angle is larger than the jet opening angle, the afterglow is characterised by a rising light-curve which peaks at a much later time than the on-axis case. The prompt emission component will not be observed due to relativistic beaming of the synchrotron photons in the ultra-relativistic jet, hence these are considered among "orphan" afterglows. In this case, the late-time (~days) X-ray emission may reveal the presence of the off-axis afterglow, as was the case with SN2020bvc (Izzo et al., 2020). The coverage of GRB-SNe in the X-ray regime is relatively good thanks to rapid-slew GRB-dedicated satellites such as *Swift* (Gehrels et al., 2004).

At radio wavelengths, a GRB afterglow can last for months to years. Radio observations can be used to determine the cooling regime of the GRB based on the behaviour of the radio lightcurves (e.g. Rhodes et al., 2020). In the case of SN2020bvc, the availability of radio to X-ray monitoring provided conclusive evidence for an off-axis GRB component (e.g. Izzo et al., 2020). It should be noted that in the majority of cases GRB-SN radio data is unavailable due to poor monitoring and/or a lack of publicly available data. Increasing the coverage of potential GRB-SNe in radio (especially SNe associated with orphan afterglows) should be a key priority for the study of GRB-SNe.

*2.2. Advantages of a webtool*

Static catalogues of GRB-SN associations have been presented in the literature prior to this work (e.g. Cano et al., 2017). However, as new GRB-SNe are discovered each year, a dynamic publication method that can be updated as new GRB-SN associations are confirmed in the literature is required. In this instance, an actively maintained webtool consisting of a database coupled to a web interface presents a simple solution to this problem. A webtool also allows missing/erroneous data to be flagged by its users, reducing the potential for propagation of mistakes through the literature (e.g. the omission of GRB071112C Klose et al., 2019a from Cano et al., 2017).

Webtools have already found applications in GRB and SN science, including: *GRBSpec*[5] (de Ugarte Postigo et al., 2014a), which provides spectra of merger and collapsar GRBs with or without associated supernovae; *GRBLC*[6] (Dainotti et al., 2024), which provides the optical/infrared lightcurves of merger and collapsar GRBs with or without associated supernovae; the *Transient Name Server*, used for reporting new optical transients including supernovae; the *Weizmann Interactive Supernova Data Repository*[7] (WISeREP) (Yaron and Gal-Yam, 2012), which hosts the spectroscopic and photometric data for some supernovae; and the *Open Supernova Catalog*[8] (Guillochon et al., 2017) which contains detailed photometry and spectroscopy for some supernovae. Thus, prior to this work, no single tool had been developed to meet the critical need of astronomers studying GRB-SN associations: a multiwavelength view of every event.

A webtool, such as the one presented here, has the capability of providing a wide range of machine-readable data in an easy to access format. Besides the capability of combining data from different sources, a webtool can also provide useful additional information about the sources (such as $E_{iso}$, $E_p$, $E_k$ etc.), which typically are only available in publications on NASA ADS.[9] Gathering this information is time consuming, and results in duplication of effort by researchers studying GRB-SNe. A webtool allows the data to be gathered in one place (with links to all primary sources), while offering the possibility for researchers to upload their new measurements as they are discovered or updated. Of course, a downside of a webtool is the need to rely on scientists to communicate their new results to keep the webtool up to date. The existence of a Python Application Programming Interface (API) for NASA ADS makes it easier to automate the search for new updates both in data and new measurements.

A significant advantage of a webtool over static databases is the possibility of fully interactive figures, which users may generate and download.

Whilst gathering the data for this tool, it became apparent that there is a major issue with data availability and standardisation in the astronomical literature. While some papers make their data available in machine readable format, many still fail to do this. In cases where the data is available, there is no consistent formatting in widespread usage. For example, some of the text files provided in papers contain bespoke, overly verbose headers which must be handled manually, slowing down the data ingestion process, particularly for large, heterogeneous datasets. Additionally, although some publications provide text files for their data they often neglect to include header lines making it impossible to determine what each row contains, or which units to use with a particular row.

Further to the issues with data formatting, there appears to be no consensus among astronomers on units or the presentation of units within papers. Compounding this is the fact that each publication presents the data slightly differently. For optical data in particular the user may need to separate the measurements from their associated uncertainty, with the machine readable table having been optimised for creation of the in-text table. Absence of units in the table, is commonplace optical magnitudes. In situations like this the appropriate units must be found elsewhere in the published text, or in some cases are missing entirely. In some cases, the data from different papers for the same event will use incompatible units. For example in radio astronomy, the observations may be reported in terms of flux (a measure of the radiation received by the detector — measured in Janskys (Jy)) or as brightness (a measure of the flux per unit solid angle — often measured in Jy/beam, with the beam being analogous to the field of view in optical astronomy). Since these units measure slightly different quantities for the source, it is necessary to convert between them when plotting the radio data for a GRB-SN.

All of these issues greatly reduce the speed at which the data can be gathered, and increase the likelihood that mistakes will be made during the analysis. By bringing many data streams together into one webtool, and reducing them so that they are all of a standard format, these pitfalls may be avoided. In the case of a webtool such as this one, since all the data available remains fully traceable, the users can scrutinise any data for themselves before including it in a publication.

**3. The GRBSN webtool**

The GRBSN webtool[10] is a public facing web application that hosts the most complete list of GRB-SN associations collected to date. The GRBSN webtool can be accessed at https://grbsn.watchertelescope.ie/. The webtool is implemented using the Python Flask[11] framework, which facilitates construction of web applications using Python and HTML. The tool allows users to view, interact with and download the data associated with any GRB-SN. Fig. 4 shows the *Home* page of the GRBSN webtool.

*3.1. Individual GRB-SN pages*

Fig. 5 shows what a user sees when navigating to one of the dedicated GRB-SN *Event* pages. The metadata are presented in two tables at the top of the page, one for the GRB and one for the SN. This data is pulled in from a Structured Query Language (SQL) database, and can be downloaded as a text file using the links on this page. The tables also contain hyperlinks to view the sources of the data.

---

[5] http://grbspec.eu/
[6] https://grblc-catalog.streamlit.app/
[7] https://www.wiserep.org/
[8] The web interface is no-longer available, however the data may be obtained at https://github.com/astrocatalogs/supernovae.
[9] https://ui.adsabs.harvard.edu
[10] https://grbsn.watchertelescope.ie/
[11] https://flask.palletsprojects.com/en/2.0.x/#api-reference

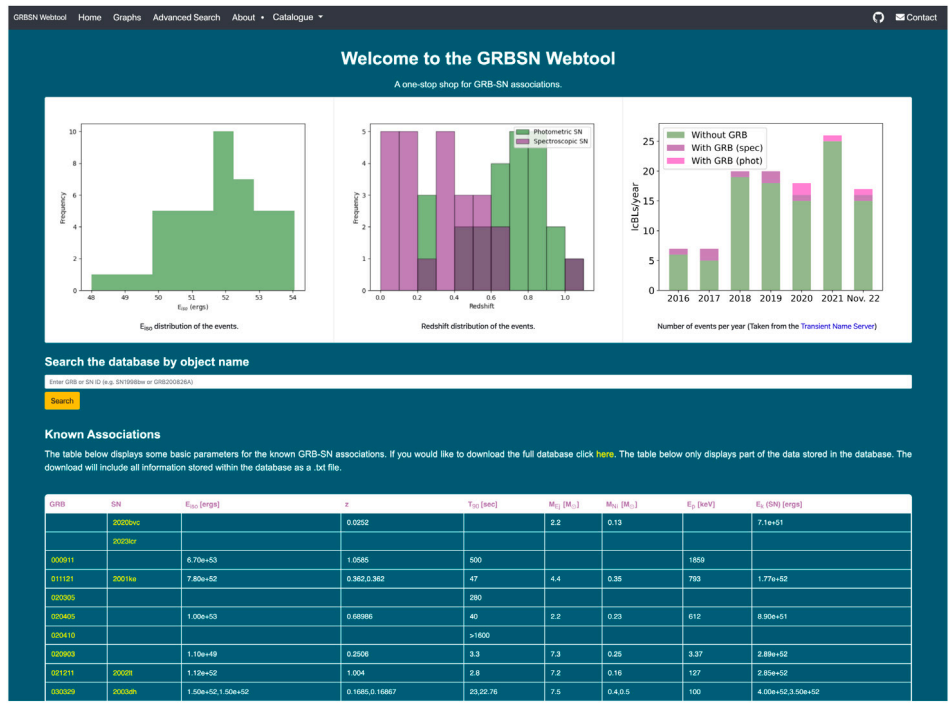

**Fig. 4.** The *Home* page of the GRBSN webtool, available at https://grbsn.watchertelescope.ie/.

These pages also display the multi-wavelength lightcurves for each GRB-SN. These graphs are created using the Bokeh[12] framework and are fully interactive (zoom, pan, save etc.). All data shown in the graphs, as well as any data not yet fully standardised, can be downloaded from this page. If there is no data for a particular wavelength, a 'No data' notice will appear on the relevant plot. The *Event* pages also feature a table of citations for all data contained in the plots and tables. These citations are provided to the user whenever data is downloaded from the webtool.

### 3.2. Graphing tool

Fig. 6 shows the *Graphs* page of the webtool,[13] which allows the user to generate custom figures from some of the parameters saved in the database. First they must choose a class of events from the following list: All Events; Orphan GRBs only (those with no prompt emission); Spectroscopic (those with an SN confirmed with spectroscopic data); or Photometric (those with no spectroscopic data). They then select the parameters they would like to plot and click plot. The tool then pulls the requested data from the database and returns a Bokeh plot of the data which the user can download as an image or as a txt file containing the data.

### 3.3. Search tool

The *Advanced Search* page, shown in Fig. 7, allows the user to query the database based on a selected subset of parameters. Users can enter min and max values for any of these parameters or a combination of these parameters. The tool then creates an SQL query and serves a table matching the users selection. This table can then be downloaded, along with the observations files for the GRBs matching the query.

### 3.4. Github

The future of the GRBSN webtool is envisioned as a community-driven project. To this end, the source code and data are fully open source, and are hosted in a Github repository.[14] Documentation including details on how to standardise and upload data to the webtool are also presented on GitHub. With the project hosted on Github it will remain available to the community even if development by the original authors ceases. This approach makes it easy for the community to suggest changes to the source code in order to add features they may need. Additionally, if users have data they would like to contribute to the project, or wish to correct an error or issue with the data, they may do this through a Github pull request. Github's integrated version control abilities also allow users to track when data is changed and see what updates have been made. This is especially important if referencing the data in a publication; in this case users can reference the relevant release of the webtool in their publication, which can be found on GitHub. The visibility of data within a public repository will also highlight the areas where more coverage is needed in future, for example the lack of radio data for many associations.

## 4. Data collection

### 4.1. Creating a catalogue of GRB-SN associations

The GRBSN webtool currently hosts 61 GRB-SN associations. The full list may be found in Appendix A; this list is complete up to the end of 2023. These events can be divided into 3 categories: 29 GRBs with a spectroscopically confirmed supernova ("Spectroscopic GRB-SN"), 31 GRBs with a photometrically confirmed supernova ("Photometric GRB-SN") and 1 off-axis GRB detected through its association with a Ic-BL supernova ("Orphan GRB-SN"). The majority of the GRB-SN associations in the tool have been confirmed in peer reviewed journals; the others are cited from the relevant General Coordinates Network (GCN) Circulars.[15] The catalogue was built up starting with the 48 associations listed in Cano et al. (2017). The remainder were gathered using data from the list of Ic-BLs on the Transient Name Server (TNS), GRBs from the Swift XRT Catalogue (Evans et al., 2009, 2007), GCN Circulars, and the GRBSpec (de Ugarte Postigo et al., 2014b) tool. These lists were cross-referenced by redshift and coordinates to find new associations discovered after the (Cano et al., 2017) paper. These associations were then confirmed by searching for peer-reviewed papers on NASA ADS. A search for new associations was also performed manually using NASA ADS.

---

[12] https://docs.bokeh.org/en/latest/index.html
[13] https://grbsn.watchertelescope.ie/graphing
[14] https://github.com/GabrielF98/GRBSNWebtool

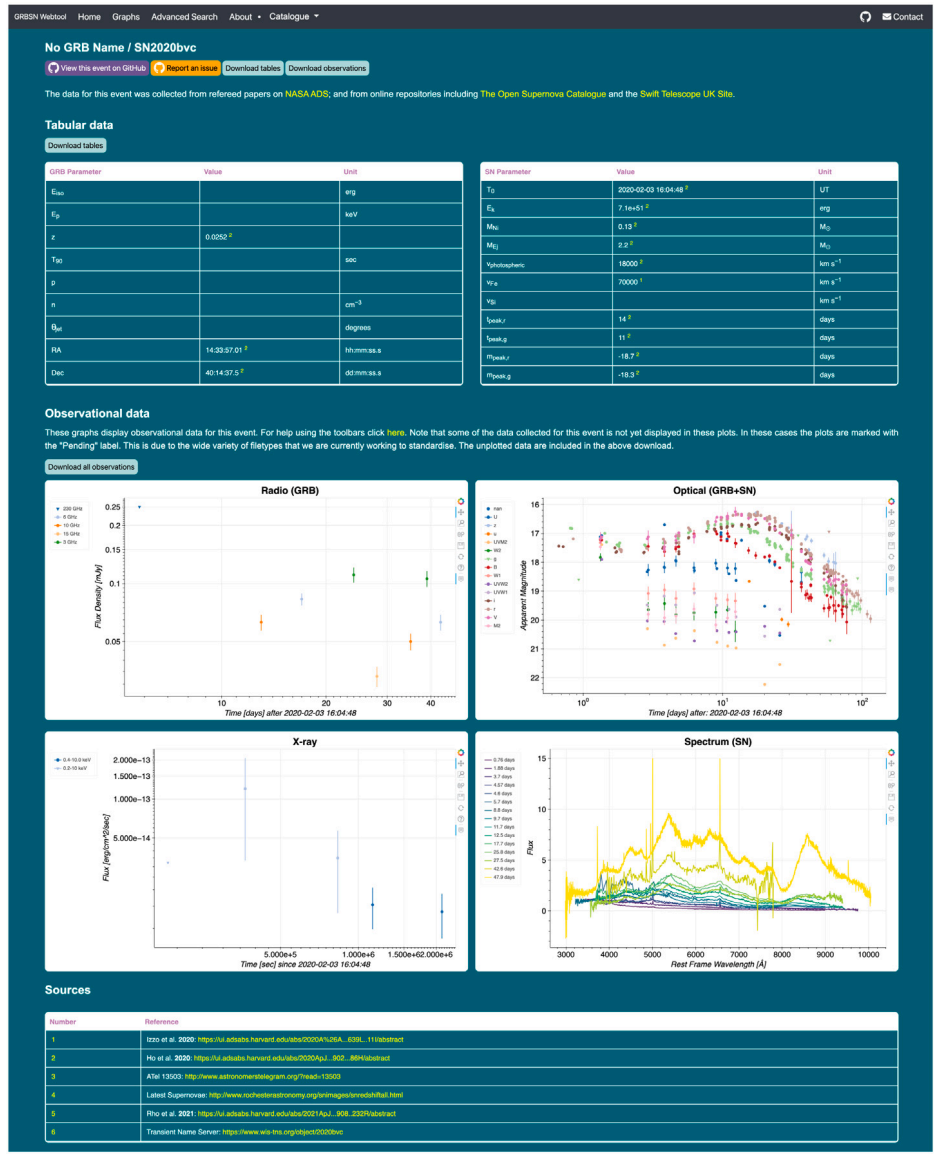

**Fig. 5.** View of the *Event* page for SN2020bvc.

Suspected GRB-SN associations may also appear in the tool, with their addition based on observations with evidence for a supernova in a GCN Circulars. To alert users to the unconfirmed nature of these events, the webtool presents a warning at the top of these pages. Additionally the database features a *notes* column which labels these as associations based solely on GCNs. This will allow users to exclude these unpublished associations from their research if they wish.

In the case of orphan GRBs, in which the Ic-BL is discovered prior to the GRB, it is possible that no GCN will be issued. In this case a new Ic-BL SN will likely be reported to TNS, which will motivate follow-up observations to search for signatures of a GRB in X-ray or Radio. Such associations may only be added to the webtool once a refereed paper appears on NASA ADS, as there can be great uncertainty around such a detection at the GCN stage. In this case the project will be reliant on the community flagging missing events on the tool. For this reason it was important to make it easy to report issues via GitHub.

In the future these two methods will determine which new associations are added to the tool. The addition of early data from GCNs or other sources will increase the visibility of new events, which may spur enhanced follow-up or collaboration between groups, prior to publication of the full set of observations. These will then be updated with refereed papers once they become available to ensure that the complete dataset is available.

---

[15] https://gcn.nasa.gov

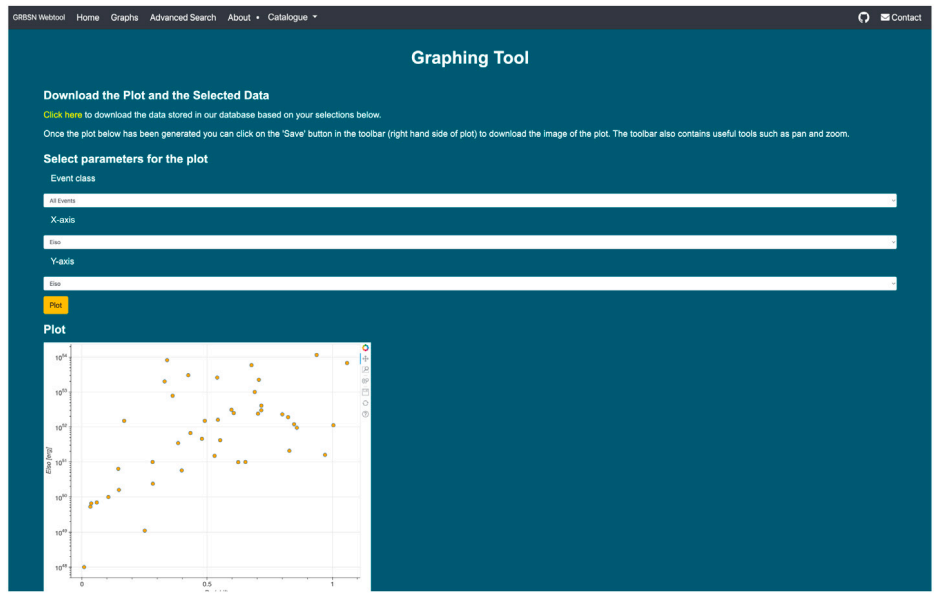

**Fig. 6.** The *Graphs* page of the GRBSN webtool, showing the integrated graphing tool.

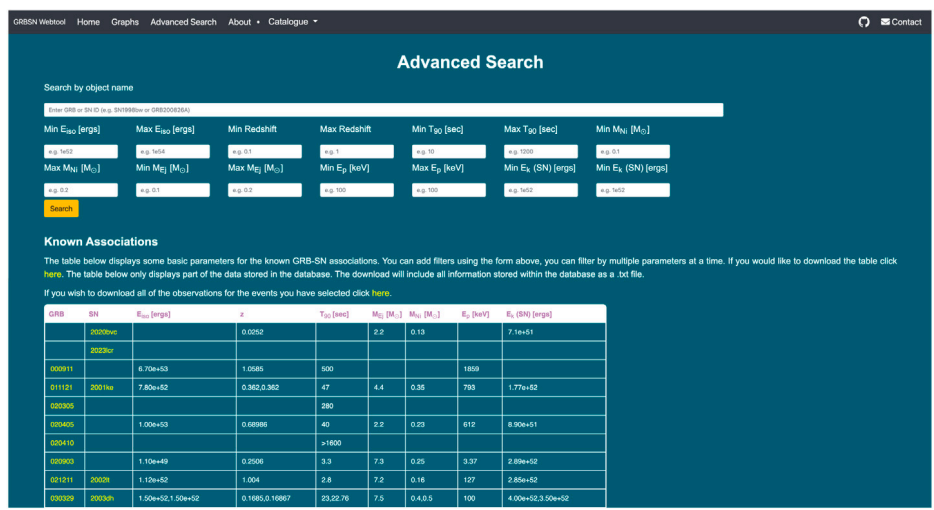

**Fig. 7.** The *Advanced Search* page of the GRBSN webtool, showing the integrated search tool.

### 4.2. Data collection and storage

The GRBSN webtool contains two types of data: observational data and event metadata. Examples of observational data includes lightcurves and spectra, while metadata refers to numeric quantities measured or calculated from these observations (redshift, isotropic energy, duration of the events etc.). Fig. 8 shows a block diagram of the webtool's primary components.

#### 4.2.1. Metadata

The metadata are stored in an SQL database, which is accessed by the webtool using the sqlite3[16] Python library. The database is split into three sub-tables. The main table contains data that was gathered from refereed papers using the NASA ADS[17] for each association. This includes measurements such as the isotropic energy, redshift, $T_{90}$, ejecta mass, nickel mass and more. The second table contains the peak magnitudes of the supernovae, both absolute and apparent, again gathered from the refereed papers found through the NASA ADS repository. The third table contains coordinate and trigger time/start time information. The data were taken from the Swift telescope website,[18] GCNs, and from refereed papers. All tables list the uncertainties of the parameters where available. It should be noted that no alteration to the quantities gathered from the literature has been made, and it should not be assumed that all quantities of a particular type were computed based on similar assumptions. If required, detailed calculations of these parameters can often be found in the relevant publications, which are linked in the SQL database.

Within the database there are multiple entries for each GRB–SN association. This arose as a natural consequence of gathering the data,

---

[16] https://docs.python.org/3/library/sqlite3.html
[17] https://ui.adsabs.harvard.edu
[18] https://swift.gsfc.nasa.gov/archive/grb_table/

**Table 1**
A sample of GRB and SN metadata parameters hosted in the GRBSN webtool database. Typical values are based on the GRBSN webtool sample.

| Parameter | Units | Explanation |
|---|---|---|
| $T_0$ | UTC | Detection time of GRB from spacecraft OR estimated explosion time of supernova obtained from lightcurve/spectrum modelling. |
| $E_{iso}$ | erg | The energy release of the GRB, based on the prompt emission and assuming isotropic emission. Typically between $10^{47}$–$10^{55}$ erg. |
| $E_p$ | keV | Peak energy of the GRB. Related to the turnover point and low-energy slope of the synchrotron spectrum. Typically between 1–1000 keV. |
| z | | Redshift. |
| $T_{90}$ | sec | Duration of the GRB prompt emission. Measured as the time taken for 90% of the flux to arrive in a GRB detector. Typically between 1–1000 s |
| p | | Electron spectral index. Describes the energy distribution of the electrons responsible for the synchrotron emission in the afterglow of the GRB. Typically 2-3. |
| n | $cm^{-3}$ | The density of the circumburst medium. Typically 0.1–1 $cm^{-3}$. |
| $\theta_{jet}$ | degrees | Half-opening angle of the GRB jet. Typically a few degrees. |
| RA | hh:mm:ss.s | The GRB right ascension. |
| Dec. | dd:mm:ss.s | The GRB declination. |
| $E_k$ | erg | The kinetic energy of the supernova ejecta. Typically $10^{50}$–$10^{52}$ erg. |
| $M_{Ni}$ | $M_\odot$ | The nickel mass within the supernova ejecta. Typically 0.1–1 $M_\odot$. |
| $M_{Ej}$ | $M_\odot$ | The supernova ejecta mass. Typically 0.1–1 $M_\odot$. Typically 1–10 $M_\odot$. |
| $v_{photospheric}$ | km s$^{-1}$ | The velocity of the supernova photosphere at peak light. |
| $v_{Fe}$ | km s$^{-1}$ | The velocity of the $v_{Fe}$ feature of the supernova at peak light. Typically 20000–25000 km/s. |
| $v_{Si}$ | km s$^{-1}$ | The velocity of the $v_{Si}$ feature of the supernova at peak light. Typically 15000–20000 km/s. |
| $t_{peak,x}$ | days | The time of peak light in band X. |
| $m_{peak,x}$ | AB/Vega | The magnitude of peak light in band X. |

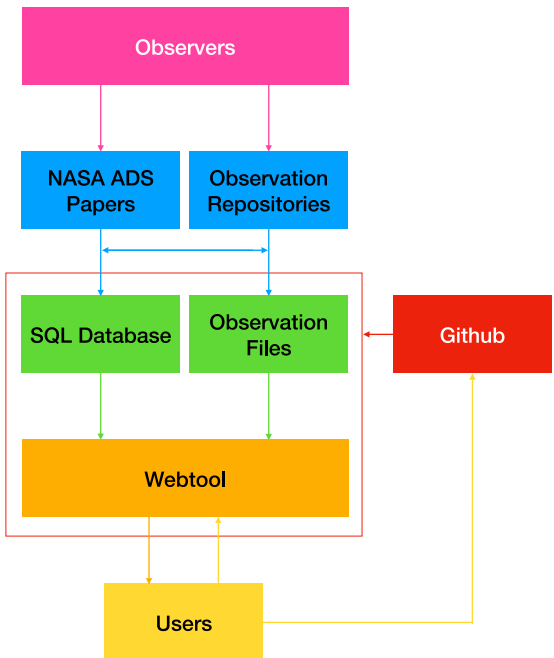

**Fig. 8.** Block diagram showing the main components of the GRBSN webtool.

as multiple papers exist for each event, each with different methods resulting in slightly different values for the same quantity. All data gathered from these papers is displayed in the webtool, in order to present a full and nuanced overview of each association. Thus, some tables in the webtool show multiple values for a parameter. Once data is downloaded it is relatively simple for the user to deal with these multiple values.

Examples of some of the parameters contained in these databases can be found in Table 1. These parameters were chosen as are they are some of the most commonly studied GRB-SN parameters appearing in the literature. Some of these parameters appear in GRB and SN emission models such as the fireball model (e.g. `afterglowpy` Ryan et al., 2020) and the Arnett model (Arnett, 1982) (e.g. MOSFiT (Modular Open-Source Fitter for Transients) Guillochon et al., 2018). In particular the SN parameters can be used to gain a detailed understanding of the strength of the SN signal detected, while the GRB parameters link the afterglow observations presented in the tool to the information contained in the GRB prompt emission. It is possible that more parameters (e.g. long/short, collapsar/merger, SN type, effective temperature of the SN, GRB jet type) may be added in future as the number of well-studied evens with detailed modelling increases, and there is in principle no restriction on users contributing new metadata to the wetbool. Detailed instructions on how users can submit their own metadata to the GRBSN webtool may be found in Appendix B.

*4.2.2. Observational data*

The GRB-SN observations are organised into folders for each event. These folders contain observations from a variety of sources, and are labelled by the GRB and SN names (e.g. GRB980425-SN1998bw). In the case of manually gathered data, these data are stored in a tab-separated text file format, which is described in the next section. For the optical spectra and photometry obtained through the Open Supernova Catalogue (Guillochon et al., 2017) API, the json format provided by this tool has been retained. Similarly, the X-ray data sourced from the Swift/XRT lightcurve repository (Evans et al., 2007, 2009) retains its own custom file format.

A text file containing the citation information for these observation files is also stored in this folder. A example file can be found in Appendix B, along with instructions on how users can submit their own data for presentation in the webtool. This text file also lists the type of observation (Optical, NIR, UV, X-ray, Radio, Spectra etc.) and the time period covered by the observation (Early: prior to day 15 of the event,

Late: Post day 15 of the event). Some of these categories were mainly used during data gathering but may be of interest to some users.

### 4.3. Data standardisation

The data gathered for the GRBSN webtool constitute a heterogeneous dataset, including data in plain text, LaTeX, JSON and HTML formats. Although all of these are common datatypes, they are not inherently interoperable, requiring different code to handle each type. These differences make it very difficult to plot and compare the datasets. To deal with this issue, the content and formatting of these files has been standardised using the following schema.

#### 4.3.1. General processing steps

The original observation files sourced from the literature were first converted into a tab-separated values format. This step had to be carried out manually due to the diversity of file types and contents. The names of these new files are determined by: the name of the GRB-SN; the data type (Optical, Radio, X-ray or Spectra); and a numerical identifier used in cases where there are multiple files of the same data type for one event. We retain the original files in the "OriginalFormats" folder, which can also be downloaded.

Following the creation of the text file, the column names were mapped to a standard schema, which is described in full on GitHub[19] and in Appendix B. There are five major column types: data columns, e.g. `mag` for magnitude; type columns, e.g. `mag_type`; unit columns, e.g. `mag_unit`; error columns, e.g. `dmag`; and boolean flag columns, e.g. `extinction_corrected`. Available column names are specific to each data type, however all files must contain the `time` column, along with at least one data column and associated columns. While column names may differ between files with different data types (Radio, Optical, X-ray, Spetra), all observational data within these types has been cast into the same format, with conversions performed where necessary. Section 4.3.2 provides further detail on the standard units used for each data type.

This file format lends itself well to use with the Pandas[20] Python module (Wes McKinney, 2010; The pandas development team, 2020). For further information on the schema used in the tab-separated files and for tutorials on using these datasets, see the GitHub Wiki.[21]

The `time` column holds the elapsed observer time since the explosion. Where data have been published with reference to the date of observation, the `time` column has been computed from the explosion epoch and the observation date, without accounting for redshift. For GRB-SNe the explosion epoch is the trigger time of the GRB, whereas for an orphan-GRB, the explosion time of the SN is used. In a future update to the tool, it will be possible to toggle rest-frame times for all events, however currently no such changes have been made to the times as provided in the source material. This is in part due to issues in confirming which times are already quoted in the rest frame.

A `README` file was created for each GRB-SN during the standardisation process to provide a link between the tab-separated files and any metadata or notes that were associated with the downloaded data or its processing. This may include notes about observing conditions and instruments used, a link to the source publication and information about the type of data contained in each observation file for the GRB-SN. This file is also viewable as markdown on GitHub.[22]

#### 4.3.2. Homogenisation of units

The standardisation of multiple inhomogeneous data sources has been a major challenge in creating this webtool. Drawing data from multiple sources has resulted in the need to standardise varying units, normalisation criteria, reference frames and scaling for the data. The following units were used for standardisation purposes: flux [erg cm$^{-1}$ sec$^{-1}$] for X-ray plots; magnitudes [AB or Vega] for optical plots; and flux density [mJy] for radio plots. These units were chosen as they are the most commonly used units for these wavelength ranges and thus will be familiar to potential users of the tool. In some cases this required the conversion of units, which was normally handled during the creation of the master files (see Section 4.3.3). In this way the original units are retained in the tab-separated files containing the original data.

In some instances, units have been omitted from tables presented in the literature. The most common example of this is the omission of units for optical magnitude. Where the magnitude information was unavailable, the units column will be populated with "unspecified", as there is no obvious method to deal with omitted units in cases like this. This of course makes it difficult to directly compared observations in the same filters taken on different instruments and to perform modelling. This difficulty emphasises the need for stricter publication guidelines in relation to reporting of units.

Optical spectroscopy is another area with a lack of standardisation, as normalisation, units and scaling factors are often different between publications and online repositories. To deal with this issue, spectra plotted in the webtool are standardised so that their flux at rest-wavelength 5000 Å is equal to 1, or the flux nearest to this wavelength if the spectrum does not include 5000 Å. The benefit of this method is that it does not rely on a complex transformation of the existing data (e.g. by the removal of the continuum) and that it allows data from multiple sources to be plotted on the same axes with similar feature sizes.

Any remaining non-standard units are clearly identified for each datapoint presented in the tab-separated files or plotted on the webtool. A brief description of the units used for each point in these plots is provided if the user hovers over a plotted point. This is intended to inform users of the issues that still exist in regard to homogenising the unit space of such a large dataset.

#### 4.3.3. Creating master files

The final step in file standardisation is to create a master file for each data type (Optical, Spectra, Radio & X-ray). This master file is generated by merging all of the standardised text files for the relevant wavelength range. This master file contains the reference information for each piece of data, allowing the user to follow the whole chain of data processing back to the source. The master files are the source of the plotted data for each event.[23] As a result, in the case of data yet to be standardised, some observational data may not appear in the webtool plots. This has been handled by means of a "Pending" notice on plots where no data appears. All of the observation files (including the master files) for each GRB-SN in the catalogue are available to download, even where not yet standardised.

### 5. An example application of the GRBSN webtool

To demonstrate the capabilities of the GRBSN webtool for rapid analysis of GRB-SN data, an investigation of the Amati relation (Amati et al., 2002) for GRB-SNe was performed and compared to that of

---

[19] https://github.com/GabrielF98/GRBSNWebtool/tree/master/Webtool/static/SourceData
[20] https://pandas.pydata.org/docs/getting_started/install.html
[21] https://github.com/GabrielF98/GRBSNWebtool/wiki
[22] https://github.com/GabrielF98/GRBSNWebtool/tree/master/Webtool/static/SourceData/GRB000911

[23] With the exception of some optical and spectral data which is plotted from the Open Supernova Catalogue JSON format, and any of the *Swift* X-ray data which is plotted from the downloaded text files.

**Table 2**
Amati relation parameter estimates.

| Events | K | m | Source |
|---|---|---|---|
| 33 | 164 ± 15 | 0.645 ± 0.029 | This work - GRB-SNe |
| 95 | 93 ± 12 | 0.604 ± 0.042 | This work - GRBs only |
| 29 | 90 ± 8 | 0.496 ± 0.037 | 1 |
| 70 | 94 ± 2 | 0.57 ± 0.01 | 2 |
| 61 | 101.05 ± 0.18 | 0.566251 ± 0.000016 | 3 |

1. Friedman and Bloom (2005); 2. Amati et al. (2008); 3. Wang et al. (2016)

GRBs without associated SNe. The Amati relation for GRBs is described as (Amati et al., 2008; Amati, 2006):

$$\frac{E_{p,i}}{1\,\text{keV}} = \frac{K E_{iso}^m}{1 \times 10^{52}\,\text{ergs}}, \quad (1)$$

where $E_{p,i}$ is the intrinsic (rest-frame) peak energy of the prompt emission that can be derived from the spectrum of the GRB and its redshift, and $E^{iso}$ is the isotropic gamma-ray energy, a measure of the total energy emitted by the GRB assuming that it is emitted isotropically (e.g. Zhang, 2018).

We used the GRBSN webtool's search tool (see Section 3.3) to download all metadata associated with the full sample of GRB-SNe in the tool. We did not impose any cuts on the data at this stage. Following this any events that had values for $E_{iso}$, $E_k$ and redshift were selected for this analysis. Where multiple values were available for one event, we selected the first value in the database, unless it was a limit value or lacked an associated uncertainty, in which case we selected the next available value for that quantity.. Limit values were only selected where there was no alternative. In total 33 events had values for all three quantities, 3 had upper or lower limits and one event had no uncertainty on $E_{iso}$. Table C.4 presents all of the events taken from the GRBSN webtool.

Although the GRBSN webtool can accommodate asymmetric uncertainties, due to the diverse origin of the measurements and the test nature of this demonstration, symmetric uncertainties were considered for $E_{iso}$ (there were no cases of assymetric uncertainties for $E_p$). The uncertainties for $E_{iso}$ were selected from the negative uncertainty column in the downloaded webtool database for simplicity. In many cases, the uncertainties reported in the literature are symmetric, so the effect on the results of our fits should be marginal.

We then used Eq. (1) to fit our dataset using orthogonal distance regression (ODR) implemented in the `scipy` Python library (Virtanen et al., 2020). This method was selected as it can account for the large uncertainties on the peak and isotropic energies simultaneously. The fit is shown in Fig. 9, with the best fit parameters for Eq. (1) included in Table 2.

To compare this relation to that of GRBs without SNe, a comparative sample was generated by combining data taken from Amati et al. (2008) and Wang et al. (2016). This sample was filtered to remove any GRBs with associated SNe, based on those in the GRBSN webtool. We then fit this combined sample using an identical ODR method to that described in the preceding paragraph.

GRB-SNe show a steeper correlation at the three sigma level compared with GRBs without SNe, while the difference in the K parameter of the fits is significant at greater than two sigma. At low isotropic energies the GRB-SN data is sparse but appears to lie above the fit. This effect is not reflected by the fit but is a known property of low luminosity GRBs (e.g. Dereli et al., 2015). As the data for typical GRBs without SNe does not extend to these low isotropic energies it is not clear whether this is an intrinsic property of GRB-SNe or of low luminosity GRBs more generally.

This ease with which data was gathered from the GRBSN webtool for this analysis clearly demonstrates the benefits of this tool. This is especially clear when compared to the data gathering process for data from (Wang et al., 2016; Amati et al., 2008), which required additional processing steps. It also demonstrates the contribution of the GRBSN webtool to furthering reproducibility goals, by making the full dataset available to all researchers via a common interface. The existence of the webtool facilitates access to this data in a way that had not existed before. By offering all these data in one place, global studies of the GRB-SN population and its properties can be done.

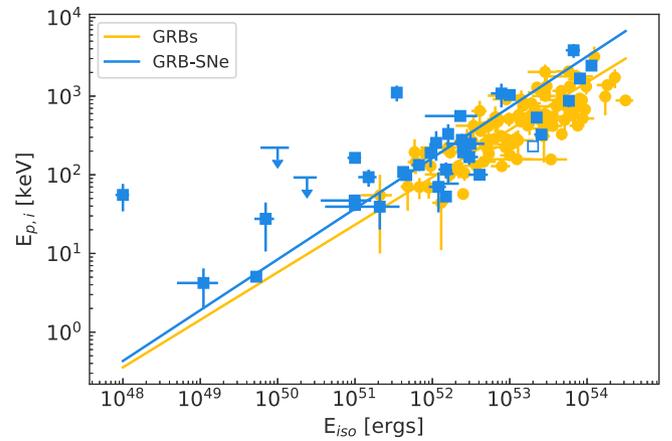

**Fig. 9.** Comparison of the Amati relation for GRBs with and without an associated supernova. Yellow points are GRBs without an associated SN (Amati et al., 2008; Wang et al., 2016). Blue points are GRB-SNe taken from the webtool database, see Table C.4. Limits (arrows) and points with no uncertainty estimate available (unfilled squares) were excluded from the fitting and are plotted only for completeness.

## 6. Future plans

### 6.1. Data visualisation

To enhance the utility of the webtool, a comparison tool will be incorporated in a future update. This tool will allow a user to select several GRB-SN events and directly compare their bulk parameters and observations. In particular it would be beneficial to be able to compare the lightcurves and spectra for any GRB-SNe within the tool. A tool which will allow for direct measurement of spectroscopic velocities of supernovae is also planned.

### 6.2. Standardisation

The standardisation of the data led to the creation of "Master" text files for Optical, Spectroscopic, Radio and X-ray data for each GRB-SN, containing all of the data for a particular GRB-SN. Tutorials for using this file format can be found on the GRBSN webtool Github. Users of the webtool can download and make use of these programs with basic software and with minimal adjustment of that software when plotting different observation types (radio, optical, X-ray etc.). Enhancement of these tutorials and expansion of their scope will enhance the user experience in future.

Eventually, we plan to convert the existing GRBSN webtool master text files to Flexible Image Transport System (FITS) table files. At present, each master text file contains all of the peer-reviewed, publicly available, fully reduced observation data associated with a GRB-SN for a particular data type (e.g. Radio, Optical, Spectra, X-ray). The use of FITS tables will allow us to group these multiwavelength observations for a particular association together in a single file, by taking advantage of the option for multiple tables in each FITS file. This will make it more straightforward for the user to handle the large amount of data that exists for many of these events. These files will be used by the tool for the plotting of observations and will be made available to users through the existing download methods on the tool, as well as via an API that will be developed around the new data formats. To supplement this, there will also be a method to add fully processed and reduced observational data in the FITS table format to facilitate the addition of newly acquired data by users.

In parallel with the creation of the webtool, artificial intelligence tools have become more readily available. In future the steps described above, particularly the initial parsing of the data presented in a paper, may be handled using artificial intelligence. However, today it is not possible to perform such a step without the risk of losing data, and so this approach was not used for this project. A major improvement could be made by mandating consistent file formats during the review phase prior to publication, with emphasis on including all relevant information, including units in all files.

*6.3. Maintaining the catalogue*

In order to keep track of new associations, the GCN notices for GRBs and the TNS will be monitored for confirmation of potential GRB-SN associations. If there is evidence of an association then it will be added to the tool. Data will then be sourced from the publicly available papers on NASA ADS when they become available. In this way the webtool will be able to keep up with the majority of new associations. The webtool benefits from being self-correcting, as missing associations are likely to be noticed and flagged by the community. For this reason, contact details are provided on the *About* page of the tool. In this manner, missing historic events and new GRB-SNe may be detected. If new evidence is presented for an old event that shows it is an association then it can be included as well. Hosting the webtool on Github also allows users to log any issues related to the tool itself or its content. Several template issues are available to allow users to flag a missing event, report issues with the data or to contribute their own data. The ability to contribute new data will be especially important as the dataset grows rapidly in the era of deeper large sky surveys.

*6.4. API*

In recent years, API access has become an essential part of any online repository. An API[24] is currently under development for the webtool. This will allow users to access data programatically for any GRB-SN association, which is immensely useful when performing analysis on a large dataset. The API is implemented using the Flask-RESTX[25] extension, and interaction is via a Python script available on GitHub. The API currently allows a user to download all observational data from the webtool. Future development of the API will include ability to download citation information for GRB-SNe as well as metadata.

# 7. Conclusion

The field of GRB-SN science is advancing rapidly. In the coming years large scale surveys of the night sky will dramatically increase the number of known GRB-SN associations. The current landscape where data is held in telescope archives, dedicated repositories for supernovae and GRBs, and in published papers is not well suited to large scale studies of the GRB-SN population. Additionally, existing repositories focus on the SN or the GRB alone, and are not targeted at astronomers studying the links between these events. To handle the predicted influx of data and meet the requirements of GRB-SN science, a centralised repository is needed.

The GRBSN Webtool is a public facing repository for the distribution of data relating to these events. The webtool contains the most up to date list of GRB-SN associations, drawing data from multiple sources, including observations presented in peer-reviewed publications, telescope archives and other online catalogues. The selection criteria for this data are based on the particular needs of astronomers studying the GRB-SN connection. Collating the data in this way drastically reduces the time spent searching for data and allows users to focus on their scientific objectives. The main features of the tool are the ability to visualise the bulk parameters and observations for any GRB-SN, interactively plot the data of all events in the tool and the ability to download data for every association. The investigation of the Amati relation demonstrates the benefits of a centralised repository for GRB-SN data. The ability to gather all of the data at once significantly reduced the time taken to study this relation for GRB-SNe and allowed for straightforward comparison of GRBs with GRB-SNe by accurately filtering the data.

The future of the project will be driven by the needs of the GRB-SN community. To accomplish this all of the source code is publicly available on GitHub, as is all of the data collected for the associations. Users of the tool can get in touch directly or make use of the issue tracking and forking features of the public GitHub repository. This will allow the GRBSN Webtool to remain the most comprehensive repository of GRB-SN associations for the foreseeable future.

# CRediT authorship contribution statement

**Gabriel Finneran:** Writing – original draft, Software, Methodology, Investigation, Data curation, Conceptualization. **Laura Cotter:** Data curation. **Antonio Martin-Carrillo:** Writing – review & editing, Supervision, Project administration, Conceptualization.

# Declaration of competing interest

The authors declare that they have no known competing financial interests or personal relationships that could have appeared to influence the work reported in this paper.

# Acknowledgements

We would like to thank the anonymous reviewers for their time and dedication to reviewing our manuscript. GF and AMC acknowledge support from the UCD Ad Astra programme. LC and AMC acknowledge support from the Irish Research Council Postgraduate Scholarship No. GOIPG/2022/1008. For data download this research has made use of the NASA Astrophysics Data System, including the NASA ADS API; The Open Supernova Catalogue (Guillochon et al., 2017) and the Open Astronomy Catalog API v1.0; the Swift/XRT lightcurve repository (Evans et al., 2009, 2007); the Transient Name Server[26]; the GRBspec database (de Ugarte Postigo et al., 2014c).[27] The webtool backend makes use of the Python Flask framework (v1.1.2),[28]; the Bokeh[29] (Team, 2022) (v2.4.2); Astropy[30] (v5.0), a community-developed core Python package for Astronomy (Astropy Collaboration et al., 2013; Astropy Collaboration et al., 2018); Pandas (v1.2.4) (Wes McKinney, 2010; The pandas development team, 2020); Matplotlib (Hunter, 2007); Numpy (v1.22.1) (Harris et al., 2020).

---

[24] https://github.com/GabrielF98/GRBSNWebtool/wiki/API
[25] https://flask-restx.readthedocs.io/en/latest/index.html
[26] https://www.wis-tns.org
[27] grbspec.eu
[28] https://flask.palletsprojects.com/en/2.0.x/#api-reference
[29] https://docs.bokeh.org
[30] http://www.astropy.org

**Appendix A. Table of GRB-SN associations catalogued in the GRBSN webtool**

See Table A.3.

**Table A.3**
Listing of GRB-SN associations hosted on the GRBSN webtool.

| GRB | SN | $E_{\gamma,iso}$ [$10^{52}$ ergs] | $E_p$ [keV] | z | $T_{90}$ [sec] | $E_k$ [$10^{52}$ ergs] | $M_{Ni}$ [$M_\odot$] | $M_{ej}$ [$M_\odot$] | Association type | References |
|---|---|---|---|---|---|---|---|---|---|---|
| 970228 | | 1.6 | 195 | 0.695 | 56 | | | | Photometric | R99; G00; C17 |
| 980326 | | 0.48 | 935 | | | | | | Photometric | C17 |
| 980425 | 1998bw | 0.0001, 0.00009 | 55 | 0.0085, 0.00867 | 30, 18 | 5, 2.5 | 0.43, 0.45 | 8 | Spectroscopic | C17; A19 |
| 990712 | | 0.67 | 93 | 0.4331 | 19 | 2.61 | 0.14 | 6.6 | Photometric | C17 |
| 991208 | | 22.3 | 313 | 0.7063 | 60 | 3.87 | 0.96 | 9.7 | Photometric | C17 |
| 000911 | | 67 | 1859 | 1.0585 | 500 | | | | Photometric | C17 |
| 011121 | 2001ke | 7.8 | 793 | 0.362 | 47 | 1.77 | 0.35 | 4.4 | Spectroscopic | C17; G03 |
| 020305 | | | | | 280 | | | | Photometric | G05 |
| 020405 | | 10 | 612 | 0.68986 | 40 | 0.89 | 0.23 | 2.2 | Photometric | C17 |
| 020410 | | | | | >1600 | | | | Photometric | C17 |
| 020903 | | 0.0011 | 3.37 | 0.2506 | 3.3 | 2.89 | 0.25 | 7.3 | Photometric | C17 |
| 021211 | 2002lt | 1.12 | 127 | 1.004 | 2.8 | 2.85 | 0.16 | 7.2 | Spectroscopic | C17 |
| 030329 | 2003dh | 1.5 | 100 | 0.1685, 0.16867 | 23, 22.76 | 4, 3.5 | 0.4, 0.5 | 7.5 | Spectroscopic | C17; A19 |
| 030723 | | | | <0.0023 | | | | | Photometric | C17 |
| 030725 | | | | | >160 | | | | Photometric | P05 |
| 031203 | 2003lw | 0.01, 0.0086 | <200 | 0.1055, 0.10536 | 40, 37 | 6 | 0.6, 0.55 | 13 | Spectroscopic | C17; A19 |
| 040924 | | 0.95 | 102 | 0.858 | 2.39 | | | | Photometric | C17 |
| 041006 | | 3 | 98 | 0.716 | 18 | 7.64 | 0.69 | 19.2 | Photometric | C17 |
| 050416A | | 0.1 | 25.1 | 0.6528 | 2.4 | | | | Photometric | C17 |
| 050525A | 2005nc | 2.5 | 127 | 0.606 | 8.8, 8.84 | 1.89 | 0.24 | 4.8 | Spectroscopic | D06; C17 |
| 050824 | | 0.209 | 21.5 | 0.8281 | 25 | 0.57 | 0.26 | 1.4 | Photometric | C17 |
| 060218 | 2006aj | 0.0053 | 4.9 | 0.0335, 0.03342 | 2000, 2100 | 0.2, 0.1 | 0.2 | 2 | Spectroscopic | C17; A19 |
| 060729 | | 1.6 | >50 | 0.5428 | 115 | 2.44 | 0.36 | 6.1 | Photometric | C17 |
| 060904B | | 2.4 | 163 | 0.7029 | 192 | 0.99 | 0.12 | 2.5 | Photometric | C17 |
| 070419A | | 0.16 | | 0.9705 | 116 | | | | Photometric | C17 |
| 071112C | | 1.91 | | 0.823 | 15 | | | | Photometric | K19 |
| 080319B | | 114 | 1261 | 0.9371 | 124.86 | 2.27 | 0.86 | 5.7 | Photometric | C17 |
| 081007 | 2008hw | 0.15 | 61 | 0.5295 | 10, 9.01 | 1.9 | 0.39 | 2.3 | Spectroscopic | C17; O15 |
| 090618 | | 25.7 | 211 | 0.54 | 113.34 | 3.65 | 0.37 | 9.2 | Photometric | C17 |
| 091127 | 2009nz | 1.5 | 35.5 | 0.49, 0.49044 | 7.1, 7.42 | 1.1, 1.35 | 0.5, 0.33 | 2.4, 4.7 | Spectroscopic | C17; O15 |
| 100316D | 2010bh | 0.007, >0.0059 | 26 | 0.059, 0.0592 | >1300, 1300 | 1, 1.54 | 0.12 | 2.5 | Spectroscopic | C17; A19 |
| 100418A | | 0.099 | 29 | 0.6239 | 8 | | | | Photometric | C17 |
| 101219B | 2010ma | 0.42 | 70 | 0.55185 | 34, 51 | 1 | 0.43 | 1.3 | Spectroscopic | C17; O15 |
| 101225A | | 1.2 | 38 | 0.847 | 7000 | 3.2 | 0.41 | 8.1 | Photometric | C17 |
| 111209A | 2011kl | 58.2 | 520 | 0.677, 0.67702 | 25000, 10000 | 3.43, 5.5 | 2.27 | 6.79, 4 | Spectroscopic | G13; KA19; G13 |
| 111211A | | | | 0.478 | | | | | Photometric | C17 |
| 111228A | | 4.07, 4.2 | 58.4 | 0.7163, 0.71627 | 101.2 | | | | Photometric | K19; C17 |
| 120422A | 2012bz | 0.024 | <72 | 0.283, 0.28253 | 5, 5.4 | 4, 2.55 | 0.3, 0.57 | 6.1 | Spectroscopic | C17; A19 |
| 120714B | 2012eb | 0.0575, 0.0594 | 101.4 | 0.3984 | 159 | | | | Spectroscopic | K19; C17 |
| 120729A | | 2.3 | 310.6 | 0.8 | 71.5 | | 0.42 | | Photometric | C17; C14 |
| 130215A | 2013ez | 3.1 | 155 | 0.597 | 65.7 | | 0.275 | | Spectroscopic | C17; C14 |
| 130427A | 2013cq | 81, 96.1 | 1250, 1028 | 0.3399 | 160, 163 | 6.4, 6.39 | 0.4, 0.28 | 6.27, 6.3 | Spectroscopic | DP16; M14; X13; C17; A19 |
| 130702A | 2013dx | 0.064 | | 0.145 | 59, 58.881 | 0.82 | 0.37 | 3.1 | Spectroscopic | T16; C17; A19 |
| 130831A | 2013fu | 0.46, 0.724 | 67 | 0.479, 0.4791 | 32.5 | 1.87 | 0.3 | 4.7 | Spectroscopic | C14; C17; K19 |
| 140606B | | 0.347 | 801 | 0.384 | 22.78 | 1.9 | 0.42 | 4.8 | Photometric | C17 |
| 150518A | | | | 0.256 | | | | | Photometric | C17 |
| 150818A | | 0.1 | 128 | 0.282 | 123.3 | | | | Photometric | C17 |
| 161219B | 2016jca | 0.016 | | 0.1475 | 7 | 4 | 0.27 | 6.5 | Spectroscopic | A19 |
| 171010A | 2017htp | 20 | 230 | 0.33 | 160 | 0.81 | 0.33 | 4.1 | Spectroscopic | M19 |
| 171205A | 2017iuk | 0.0066 | | 0.037 | 189.4 | 0.14, 0.09 | | 1.1 | Spectroscopic | W18 |
| 180728A | 2018fip | | | | | | | | Spectroscopic | W19 |
| 190114C | 2019jrj | 30 | | 0.4245 | 361.5, 362 | | | | Spectroscopic | M21 |
| 190829A | AT2019oyw | | | 0.0785 | 58.2 | 1.36 | 0.5 | 5.67 | Spectroscopic | H21 |
| | 2020bvc | | | 0.0252 | | 0.71 | 0.13 | 2.2 | Orphan Afterglow | I20; H20 |
| 200826A | | | | 0.71 | 0.65 | 6 | | | Photometric | A21 |
| 201015A | | | | | | | | | Photometric | GCN 29033 |
| 211023A | | | | | | | | | Photometric | GCN 31596 |
| 220219B | | | | | | | | | Photometric | GCN 31739 |
| 221009A | 2022xiw | | | 0.151 | | 5.8 | 1 | 7.1 | Spectroscopic | F23 |
| 230812B | 2023pel | | | | | | | | Spectroscopic | GCN 34597 |
| | 2023lcr | | | | | | | | Spectroscopic | GCN 34385 |

References: R99: (Reichart, 1999); G00: (Galama et al., 2000); C17: (Cano et al., 2017); A19: (Ashall et al., 2019); G03: (Garnavich et al., 2003); G05: (Gorosabel et al., 2005); P05: (Pugliese et al., 2005); D06: (Della Valle et al., 2006); K19: (Klose et al., 2019b); O15: (Olivares E. et al., 2015); G13: (Gendre et al., 2013); KA19: (Kann et al., 2019); C14: (Cano et al., 2014); DP16: (De Pasquale et al., 2016); M14: (Melandri et al., 2014); X13: (Xu et al., 2013); T16: (Toy et al., 2016); M19: (Melandri et al., 2019); W18: (Wang et al., 2018); W19: (Wang et al., 2019); M21: (Melandri et al., 2022); H21: (Hu et al., 2021); I20: (Izzo et al., 2020); H20: (Ho et al., 2020); A21: (Ahumada et al., 2021); F23: (Fulton et al., 2023)

**Appendix B. Contributing data to the GRBSN webtool - all new**

This appendix provides a guide to flagging issues with the GRBSN webtool, and for contributing your own data to the tool. All submissions and issues are reviewed by the GRBSN webtool administrators at University College Dublin.

*B.1. Submitting data via the GitHub interface*

The GRBSN team use GitHub to manage the GRBSN webtool. If you wish to flag an issue or contribute data, you will need to have a GitHub account. You can register for an account here.[31] GitHub is free to use for this use-case.

If you wish to identify a missing GRB-SN or missing/erroneous data in the webtool, but you do not have any data to submit, you should use the issue tracker[32] to create an issue. You may follow the instructions in the existing template issues or start from a blank issue.

If instead you wish to submit data to the webtool, you should begin by forking[33] the repository. This will give you a replica of the GRBSN webtool repository, into which you may add your data. Please note that you must submit one pull-request per GRB-SN for which you are submitting data. Do not group multiple bursts into one pull-request.

Once you have contributed your data to your forked repository, you should submit your changes to the main repository for review using a pull-request.[34] The data you submit via pull-request must be formatted as described in Appendices B.2 and B.3.

To assist with the review of your data, your pull-request should include a text description of each file submitted, including a general overview of the data in the file and a link to the source of the data. You can also include notes to the reviewers at this stage. This issue will subsequently be reviewed by the GRBSN team prior to the addition of the data to the tool. The GRBSN team can be contacted at any time to assist with this process.

*B.2. Contributing GRB-SN observational data*

This section is relevant if you wish to submit reduced, calibrated afterglow observations to the GRBSN webtool via a pull-request.

Your pull-request for should contain: one or more *observation data files* (containing the afterglow observation data) and a YAML file called an *observation information file* (that identifies the contents of each file). These files are described in the following sections.

*B.2.1. Observation data files*

All observation files submitted to the GRBSN webtool should be in tab-separated values (TSV) format. This should consist of a header line and several data lines, as shown below:

```
1  date	time	freq	flux_density	dflux_density	date_unit	time_unit	freq_unit	flux_density_unit
2  2004-Dec-09	621.2	1.4	650	70	yyyy-month-dd	days	GHz	microJy
3  2004-Dec-09	621.2	8.5	250	30	yyyy-month-dd	days	GHz	microJy
4  2004-Dec-23	635.1	1.4	590	70	yyyy-month-dd	days	GHz	microJy
```

A detailed breakdown of potential column names that can appear in the headers of these files is given in Appendix B.2.2

When naming observational data files, the filename should contain the name of the GRB, name of the SN and the type of data in the file, for example:

– *GRB030329-SN2003dh_Radio.txt*

All filenames should contain one of the following tags:

– `Xray`: A file containing observations from an X-ray telescope/satellite.
– `Optical`: Any file containing observations in the Optical bands. `NIR`, `IR`, Visible or `UV` range. If a file contains `NIR`, `IR` or `UV` data then it should also contain the relevant tag in it's filename.
– `Radio`: Any file containing observations at Radio wavelengths.
– `Spectra`: Any file containing spectroscopic observations.

In the case of a filename with multiple tags, each tag should be separated by an underscore, for example:

– *GRB030329-SN2003dh_Optical_NIR.txt*

If there are several files in the same wavelength, e.g. multiple optical files, an index should be added to each filename, for example:

– *GRB030329-SN2003dh_Optical.txt*
– *GRB030329-SN2003dh_Optical1.txt*
– *GRB030329-SN2003dh_Optical2.txt*

---

[31] https://github.com/signup
[32] https://github.com/GabrielF98/GRBSNWebtool/issues/new/choose
[33] https://docs.github.com/en/pull-requests/collaborating-with-pull-requests/working-with-forks/fork-a-repo
[34] https://docs.github.com/en/pull-requests/collaborating-with-pull-requests/proposing-changes-to-your-work-with-pull-requests/about-pull-requests

*B.2.2. Observation file headers*

One of the major aims of the GRBSN webtool is to make all of the observational data for GRB-SNe available in one location. Key to this is the conversion of all data into a common format. This format is described in this subsection.

In each observation file you are submitting, the first row (called a header) should give an indication of the data columns contained in the file. This subsection provides a list of the possible column names that may be found in these headers.

There are five major column types: data columns, e.g. `mag` for magnitude; type columns, e.g. `mag_type`; unit columns, which use the prefix 'd', e.g. `mag_unit`; uncertainty columns, e.g. `dmag`; and boolean flag columns, e.g. `extinction_corrected`.

A special boolean flag column is used to represent upper and lower limits. Limit columns are identified by the suffix '_limit'. In these columns, upper limits are represented by a 1, ordinary values by a 0 and lower limits by −1. When submitting data to the webtool, limits must be indicated using > / < in the data column for the relevant quantity. Please ensure that you include this information in your submission, as it is used to automatically create the corresponding limit column when the GRBSN team review your data.

Although it is not necessary to include all column names in all files, there are a small subset of column names that must appear in all files. The following columns (and their associated unit and uncertainty columns) are required to appear in each observation file:

– `time` and/or `date`
– `time_frame`
– A column that corresponds to the brightness of the observed radiation.
– `instrument`
– `reference`

As the GRBSN webtool is focused on afterglow and supernova data, there is no provision for prompt emission data in the schema described here.

In terms of adding afterglow data from spacecraft/observatory not already represented on the webtool, there is no issue with adding a new instrument or filter, for example. The only requirement is that you include fully calibrated data with all relevant units, limit information, instrument information, exposure times etc. If you wish to add a new column name, simply add it to your text file, following the formatting schema described here, and mention it when you upload your data via pull-request, providing justification for its use and a brief explanation.

*B.2.3. General column names*

The following column names might appear in any file. Note that this list is not necessarily exhaustive, and new column names may be added as necessary provided this is flagged when adding the data to an issue on GitHub.

**Times and dates**

**Note:** The `time` column should always appear in all files.

– `date`: The date of observation.
– `date_unit`: The unit of the date of observation. The options are:

  – `yyyy-month-deciday`: The year, month and decimal day.
  – `yyyy-mm-deciday`: The year, numerical month and decimal day.
  – `utc = yyyy-mm-ddThh:mm:ss`: UTC time format. The year, numerical month, then a T, then day and hours, min, sec.
  – `yyyy-month-dd-hh:mm:ss`: The year, month, day and hours, min, sec.
  – `yyyy-mm-deciday-deciday`: The year, numerical month and decimal day - range.
  – `yyyy-month-deciday-month-deciday`: The year, month and deciday, both are ranges.
  – `yyyy-month-deciday-deciday`: The year, month and decimal day - range.
  – `yyyy-month-dd-hh:mm`: The year, month, day and hours and minutes.
  – `yyyy-month-dd-hh:mm-hh:mm`: Epoch range with the month as a three letter code.
  – `yyyy-mm-dd-hh:mm-hh:mm`: Epoch range in standard time format
  – `yyyy-month-dd-hh.h-hh.h`: Year, month, day and decimal hour range.
  – `MJD`: Modified Julian Day.
  – `MJD-MJD`: Modified Julian Day - range.

– `time`: The elapsed time since the reference point of the data. If known, the GRB trigger time is used, otherwise the estimated explosion time of the SN from its lightcurve should be used.
– `time_unit`: Unit for the elapsed time. The options are:

  – `seconds`
  – `kiloseconds`
  – `hours`
  – `days`

– `dtime`: Uncertainty on the time. It is assumed that these will be in the same units as the `time` column.
– `time_frame`: Are the times in the rest frame of the observer or of the event. Options are:

  – `event`: For times given in the rest frame of the event.
  – `observer`: For times given in the rest frame of the observer.

*Miscellaneous*

- `integration`: The duration of the observation. The default unit is seconds. This may have been converted from an exposure column in the original file.
- `integration_unit`: The units for the integration time. Standard time units will be used.
- `reference`: This should be ideally be a NASA ADS abstract URL. This provides a direct link to the paper or resource from which the file came.
- `instrument`: The name of the telescope/instrument system used to take data.

B.2.4. *Data type specific column names*

*Spectroscopy*

- `obs_wavelength`: The observed wavelength of the observation.
- `rest_wavelength`: The rest frame wavelength of the observation. Calculated by dividing the observed wavelength by 1+z, where z is the redshift.
- `wavelength_unit`: Unit for wavelength. The options are:

    - `angstroms`
    - `nm`: nanometers.

- `flux`: The observed flux.
- `flux_unit` Unit for the flux. The options are:

    - `uncalibrated`: used in cases where the units are unknown, or when the flux is uncalibrated.
    - `calibrated`: used when some calibration has been done but the units were not provided.
    - `erg/s/cm2/A`: erg per second per square cm per angstrom. This is the preferred unit.

- `dflux`: The uncertainty on the source flux.
- `redshift`: The event redshift. If it is not known, please flag this in your submission.
- `sky_flux`: Sometimes measured when spectra are taken. It is in the same units as the flux of the source.

*X-ray*

- `flux`: The received flux of the source.
- `flux unit`: The units used for the flux of the source. The preferred unit is `erg/cm^2/sec`.
- `dflux`: The uncertainty on the source flux.
- `flux_limit`: Is the source flux an upper limit (1), not a limit (0) or a lower limit (−1). Added by the GRBSN team.
- `energy_range`: The energy range of the xray data.
- `HR`: The hardness ratio.
- `dHR`: The uncertainty on the harness ratio.

*Radio*

- `freq` The frequency of the radio band.
- `freq_unit`: The unit for the frequency of the radio band. The options are:

    - `GHz`
    - `MHz`

- `flux_density`: The flux density of the source.
- `dflux_density`: Uncertainty on the flux density.
- `flux_density_unit`: The unit for the flux density of the source. The options are:

    - `milliJy`: milli Jansky.
    - `microJy`: micro Jansky.

- `flux_density_limit`: Is the flux density an upper limit (1), not a limit (0) or a lower limit (−1). Added by the GRBSN team.
- `seeing`: The seeing. Default unit is 'arcseconds'.
- `beam`: The size of the telescope beam.
- `beam_unit`: The unit associated with the beam. The options are:

    - `arcseconds^2`

- `bandwidth`: The bandwidth of the observation.
- `bandwidth_unit`: Unit of the bandwidth. The options are:

- GHz
- MHz

- `optical_depth`: Optical depth along the line of sight to the source.
- `polarisation`: Degree of polarisation of the source.
- `system_noise_temp`: The radio system noise temperature in Kelvin.
- `VLA_Project_Code`: Used in Very Large Array data.
- `position_angle`

*Optical*

- `mag`: Magnitude of the source.
- `dmag`: Uncertainty on the magnitude. When there are asymmetric uncertainties, a second uncertainty column can be used: `dmag2`.
- `mag_unit`: The units used for the magnitude of the source. The options are:

  - `Vega`
  - `AB`
  - `unspecified`: Used when neither AB nor Vega are clearly specified in the source material.

- `mag_type`: Supplied only when data are in absolute magnitudes. Options are:

  - `apparent`
  - `absolute`

- `mag_limit`: Is the magnitude an upper limit (1), not a limit (0) or a lower limit (−1). Added by the GRBSN team.
- `seeing`: The seeing. Default unit is `arcseconds`.
- `counts`: The total counts received by a CCD or other instrument.
- `dcounts`: The uncertainty on the counts.
- `flux_density`: The flux density of the source.
- `dflux_density`: Uncertainty on the flux density.
- `flux_density_unit`: The unit for the flux density of the source. The options are:

  - `milliJy`: milli Jansky.
  - `microJy`: micro Jansky.

- `flux_density_limit`: Is the flux density an upper limit (1), not a limit (0) or a lower limit (−1). Added by the GRBSN team.
- `extinction`: The correction to the magnitude due to extinction, measured in the associated band and with the associated units.
- `extinction_corrected`: Have the data been extinction- corrected. Options are:

  - `true`
  - `false`

- `kcorr`: The k correction. Used in optical/NIR/UV This will be followed by the relevant bands being corrected between e.g. `kcorr_vs` for correction from V to STIS.
- `wavelength`: The wavelength of the observation.
- `wavelength_unit`: Unit for wavelength. The options are:

  - `angstroms`
  - `nm`: nanometres.

- `airmass`
- `band`: The filter used for the observation. A list of the bands is available on GitHub.

*B.2.5. Observational data information file*

When submitting observation data, your pull-request should contain an observation information file in YAML (yet another markup language) format. This file is used to identify the types of observation files being added to the tool. It is also used to create the README text files for each event.

Please note that this file will already exist for most GRB-SNe in the webtool (named *readme.yml* in the folder associated with each GRB), and you should only need to edit it when making your submission.

If this file does not exist (for example for an uncatalogued association), you may create it. A template for this YAML file can be found on GitHub[35].

A description of these keys is given in the following example file. Note that the keys contained in this file must not be altered.

---

[35] https://github.com/GabrielF98/GRBSNWebtool/blob/master/observation_info.yaml

```yaml
 1  event_name: GRB000911 # Name of the event (combine GRB and SN names like: GRB130702A-SN2013dx.)
 2  grb_name: GRB000911 # Name of the GRB.
 3  sn_name: null # Name of the SN (include SN or AT etc.)
 4  status: null # Used behind the scenes of the webtool. Leave null.
 5  processed: false # Used behind the scenes of the webtool. Leave false.
 6  processinfo: null # Used behind the scenes of the webtool. Leave null.
 7  xray: false # True if you are submitting X-ray data.
 8  optical: true # True if you are submitting optical data.
 9  spectra: false # True if you are submitting optical data.
10  radio: true # True if you are submitting radio data.
11  filenames: # A listing of all filenames relating to the GRB-SN.
12    GRB000911_Optical_NIR.txt: # Name of file you are submitting. Refer to section on filenaming.
13      datatype: Late, NIR, Optical
14      sourceurl: https://ui.adsabs.harvard.edu/abs/2005A%26A...438..841M /abstract # Source of the data, commonly the ADS abstract URL of the publication from
             which the observations are sourced.
15      notes: |
16        Journal of the optical and NIR photometry of the GRB 000911 afterglow. Magnitude uncertainties are at 1$\sigma $ confidence level; upper limits are at
             3$\sigma $ confidence level. The magnitudes are not corrected for Galactic interstellar extinction.
17
18        2000-Sep-16.328 & 2000-Sep-16.328 were made from spectroscopic observations.
19    GRB000911_Optical.txt:
20      datatype: Early, Optical
21      sourceurl: https://ui.adsabs.harvard.edu/abs/2001A%26A...378..996L/abstract
22      notes: |
23        Notes: This table combines tables 1, 2 and 3 from the relevant paper.
24
25        The Keck observations come from Price et al. (2001)
26
27        The observations after the VLT list come from spectroscopic measursments.
28    GRB000911_Radio.txt:
29      datatype: Radio
30      sourceurl: https://ui.adsabs.harvard.edu/abs/2001A%26A...380...81S/abstract
31      notes: |
32        No Notes.
```

*B.3. Contributing GRB-SN metadata*

If you have GRB-SN metadata to contribute to the tool, then you must do this by pull-request as described in Appendix B.1. Data should be submitted in the YAML file format described below. A template for this YAML file can be found on GitHub.[36]

As discussed in the text, the GRBSN webtool has three SQL databases under the hood. One is used for peak time and magnitude information, one for trigger time information use to compute t0 for the observation plots, and a large dataset of GRB-SN parameters. If accepted this YAML file will be used to add one new entry to each of these databases. As a result, you must create a YAML file for each GRB-SN and for each source publication from which you have collected data (i.e. if you collected data from two papers, make two YAML files).

Note that some values may be provided with a positive and negative uncertainty in this file. In a case where both values are identical, provide both values in the file.

```yaml
 1  # Table for peak time and magnitude information.
 2  PeakTimesMags:
 3    grb_id: # e.g. GRB091214A
 4    sn_name: # e.g. AT5034aj
 5    time: 15 # Time of peak light in days since t0.
 6    dtime: 2 # Uncertainty on the time of peak light.
 7    reltime: # rest-frame or observer-frame depending on which frame the time was given in.
 8    mag: 18.3 # magnitude at peak
 9    dmag: 0.1 # Uncertainty on the magnitude
10    delmag15: 1 # Magnitude decline rate over 15 days since peak.
11    ddelmag15: 0.1 # Uncertainty on the magnitude decline rate over 15 days since peak.
12    absmag: -18.5 # Peak absolute magnitude
13    dabsmag: # Uncertainty on peak absolute magnitude
14    band: R # What band is this peak light information for.
15    unit: # Unit used for magnitude - AB/Vega
16    source: # URL for the reference. Preferred as a NASA ADS URL.
17
18  SQLDataGRBSNe:
19    GRB: GRB091102 # GRB name
20    SNe: SN2040shfds # SN name
21    e_iso: 1e52 # Isotropic energy of the GRB [erg].
22    d_neg_e_iso: 1e51 # Negative uncertainty on the isotropic energy of the GRB.
23    d_pos_e_iso: 2e51 # Positive uncertainty on the isotropic energy of the GRB.
24    ni_m : 0.3 # Nickel mass ejected by the SN [solar masses].
25    d_neg_ni_m: 0.04 # Negative uncertainty on the nickel mass ejected by the SN.
26    d_pos_ni_m: 0.04 # Positive uncertainty on the nickel mass ejected by the SN.
27    ej_m: 4 # Total mass of the SN ejecta [solar masses].
28    d_neg_ej_m: 1 # Negative uncertainty on the total mass ejected by the SN.
29    d_pos_ej_m: 2 # Positive uncertainty on the total mass ejected by the SN.
30    E_p: 100 # Peak energy of the GRB [keV]
31    dE_p: 10 # Uncertainty on E_p.
32    e_k: 1e52 # Kinetic energy of the SN [erg].
33    d_neg_e_k: 1e51 # Negative uncertainty on the kinetic energy of the SN.
34    d_pos_e_k: 5e50 # Positive uncertainty on the kinetic energy of the SN.
35    open_deg: 3 # Half-opening angle of the GRB jet [degrees].
36    d_open_deg: 1 # Uncertainty on the half-opening angle of the GRB jet.
37    v_Fe_expansion: 21000 # Expansion velocity measured from the Fe II feature of the Ic-BL spectrum at peak light [km/s].
38    dv_Fe_expansion: 4000 # Uncertainty on vFe.
39    v_Si_expansion: 15000 # Expansion velocity measured from the Si II feature of the Ic-BL spectrum at peak light [km/s].
40    dv_Si_expansion: 3000 # Uncertainty on vSi.
41    v_photospheric: 20000 # Inferred supernova photosphere velocity at peak light [km/s].
42    d_neg_v_photospheric : 5000 # Negative uncertainty for photosphere velocity.
43    d_pos_v_photospheric: 1000 # Positive uncertainty for photosphere velocity.
44    z: 0.4 # Redshift of the event.
45    dz : 0.1 # Uncertainty on redshift.
46    T90: 24 # GRB duration [sec]
47    dT90 : 3 # Error on GRB duration.
48    log_e_gam: 47.3 # Beaming corrected GRB energy [erg]. Log scaled.
49    dlog_e_gam : 42 # Uncertainty on beaming corrected GRB energy.
50    log_e_radio: # GRB energy measured in the radio regime [erg]. Log scaled.
51    d_log_e_radio: # Uncertainty on GRB energy measured in the radio regime. Log scaled.
52    t_peak : 11.5 # Peak time of the SN [days]
```

---

[36] https://github.com/GabrielF98/GRBSNWebtool/blob/master/database_entry.yaml

```
53      dt_peak: 4 # Error on the peak time of the SN.
54      n: 1 # Circumburst density [cm^{-3}]
55      p: # Electron power law index.
56      dp: # Uncertainty on above.
57      PrimarySources : # Link to the source publication when they are the source of these metadata calculations.
58      SecondarySources : # Link to the source publication if they cite other publications that computed the metadata.
59      Notes: # Notes.
60      Data: # A column used internally.
61
62   TrigCoords:
63      grb_id: GRB240927D # GRB name
64      sn_name : SN1998bw # SN name
65      trigtime: # Time of the GRB trigger or SN explosion time in UTC. Formatted as YYYY-MM-DDTHH:MM:SS.
66      ra: # The right ascension of the GRB [HH:MM:SS].
67      ra_decimal: # The decimal right ascension the GRB.
68      ra_deci_err : # Uncertainty on above.
69      dec : # The declination of the GRB [dd:mm:ss.s].
70      dec_decimal : # The decimal declination of the GRB.
71      dec_deci_err: # Uncertainty on above.
72      source: # Link to the source of this information.
73      satellite: Fermi # Name of satellite that detected the GRB.
```

## Appendix C. Table of GRB-SN associations used to measure the Amati relation

See Table C.4.

**Table C.4**
Listing of GRB-SN associations from the GRBSN webtool used for the Amati relation in Section 5.

| GRB | SN | $E_{\gamma,iso}$ [$10^{52}$ ergs] | $\Delta E_{\gamma,iso}$ [$10^{52}$ ergs] | $E_p$ [keV] | $\Delta E_p$ [keV] | z | Source |
| --- | --- | --- | --- | --- | --- | --- | --- |
| 970228 |  | 1.6 | 0.1 | 195 | 64 | 0.695 | Reichart (1999) |
| 980425 | 1998bw | 0.0001 | 0.00002 | 55 | 21 | 0.009 | Ashall et al. (2019) |
| 990712 |  | 0.7 | 0.1 | 93 | 15 | 0.433 | Cano et al. (2017) |
| 991208 |  | 22 | 2 | 313 | 31 | 0.706 | Cano et al. (2017) |
| 000911 |  | 67 | 14 | 1859 | 371 | 1.058 | Cano et al. (2017) |
| 011121 | 2001ke | 8 | 2 | 793 | 265 | 0.362 | Garnavich et al. (2003) |
| 020405 |  | 10 | 0.9 | 612 | 10 | 0.690 | Cano et al. (2017) |
| 020903 |  | 0.0011 | 0.0006 | 3 | 2 | 0.251 | Cano et al. (2017) |
| 021211 | 2002lt | 1.1 | 0.1 | 127 | 52 | 1.004 | Cano et al. (2017) |
| 030329 | 2003dh | 1.5 | 0.3 | 100 | 23 | 0.169 | Ashall et al. (2019) |
| 031203 | 2003lw | 0.010 | 0.004 | <200 |  | 0.105 | Ashall et al. (2019) |
| 040924 |  | 1 | 0.1 | 102 | 35 | 0.858 | Cano et al. (2017) |
| 041006 |  | 3 | 0.9 | 98 | 20 | 0.716 | Cano et al. (2017) |
| 050416A |  | 0.1 | 0.01 | 25.1 | 4.2 | 0.653 | Cano et al. (2017) |
| 050525A | 2005nc | 2.5 | 0.4 | 127 | 10 | 0.606 | Della Valle et al. (2006) |
| 050824 |  | 0.2 | 0.2 | 22 | 11 | 0.828 | Cano et al. (2017) |
| 060218 | 2006aj | 0.0053 | 0.0003 | 4.9 | 0.3 | 0.034 | Ashall et al. (2019) |
| 060729 |  | 1.6 | 0.6 | >50 |  | 0.543 | Cano et al. (2017) |
| 060904B |  | 2.4 | 0.2 | 163 | 31 | 0.703 | Cano et al. (2017) |
| 080319B |  | 114 | 9 | 1261 | 65 | 0.937 | Cano et al. (2017) |
| 081007 | 2008hw | 0.15 | 0.04 | 61 | 15 | 0.529 | Olivares E. et al. (2015) |
| 090618 |  | 26 | 5 | 211 | 22 | 0.540 | Cano et al. (2017) |
| 091127 | 2009nz | 1.5 | 0.2 | 36 | 2 | 0.490 | Olivares E. et al. (2015) |
| 100316D | 2010bh | 0.007 | 0.002 | 26 | 16 | 0.059 | Ashall et al. (2019) |
| 100418A |  | 0.1 | 0.1 | 29 | 2 | 0.624 | Cano et al. (2017) |
| 101219B | 2010ma | 0.4 | 0.1 | 70 | 8 | 0.552 | Olivares E. et al. (2015) |
| 101225A |  | 1.2 | 0.3 | 38 | 20 | 0.847 | Cano et al. (2017) |
| 111209A | 2011kl | 58 | 7 | 520 | 89 | 0.677 | Gendre et al. (2013) |
| 111228A |  | 4 | 1 | 58 | 7 | 0.716 | Klose et al. (2019b) |
| 120422A | 2012bz | 0.024 | 0.008 | <72 |  | 0.283 | Ashall et al. (2019) |
| 120729A |  | 2 | 2 | 311 | 32 | 0.800 | Cano et al. (2014) |
| 130215A | 2013ez | 3 | 2 | 155 | 63 | 0.597 | Cano et al. (2014) |
| 130427A | 2013cq | 81 | 8 | 1250 | 50 | 0.340 | Ashall et al. (2019) |
| 130831A | 2013fu | 0.46 | 0.02 | 67 | 4 | 0.479 | Cano et al. (2014) |
| 150818A |  | 0.1 | 0.02 | 128 | 13 | 0.282 | Cano et al. (2017) |
| 171010A | 2017htp | 20 |  | 230 |  | 0.330 | Melandri et al. (2019) |

**Data availability**

All data is available from the online repository described in this paper.